\documentclass[sigconf,screen]{acmart}

\copyrightyear{2023} 
\acmYear{2023} 
\setcopyright{acmlicensed}\acmConference[ESEC/FSE '23]{Proceedings of the
31st ACM Joint European Software Engineering Conference and Symposium on
the Foundations of Software Engineering}{December 3--9, 2023}{San Francisco,
CA, USA}
\acmBooktitle{Proceedings of the 31st ACM Joint European Software
Engineering Conference and Symposium on the Foundations of Software
Engineering (ESEC/FSE '23), December 3--9, 2023, San Francisco, CA, USA}
\acmPrice{15.00}
\acmDOI{10.1145/3611643.3616329}
\acmISBN{979-8-4007-0327-0/23/12}

\received{2023-03-02}
\received[accepted]{2023-07-27}

\AtBeginDocument{%
  }

\usepackage{amsmath,amsfonts}
\usepackage{multirow}
\usepackage{algorithmic}
\usepackage{graphicx}
\usepackage{textcomp}
\usepackage{xcolor}
\usepackage{framed}
\usepackage{listings}
\usepackage{bbding} 
\usepackage{color}
\usepackage{enumitem}
\usepackage{threeparttable}
\usepackage{xspace}
\usepackage{indentfirst}
\usepackage{booktabs}
\usepackage{subfigure}
\usepackage{balance}
\usepackage{colortbl}
\usepackage{pifont}
\usepackage{comment}
\usepackage{footnote}
\usepackage{url}
\usepackage[hyphenbreaks]{breakurl}
\usepackage{hyperref}
\usepackage{cleveref}

\usepackage[ruled,lined,linesnumbered,vlined,algo2e]{algorithm2e}
\usepackage{microtype}

\newcommand{\tool}{\textsf{Iris}\xspace}

\begin{document}

\title{Automated and Context-Aware Repair of Color-Related Accessibility Issues for Android Apps}

\author{Yuxin Zhang}
\affiliation{%
  \institution{College of Intelligence and Computing, Tianjin University}
  \city{Tianjin}
  \country{China}
}
\email{yuxinzhang@tju.edu.cn}

\author{Sen Chen}
\authornote{Corresponding author}
\affiliation{%
  \institution{College of Intelligence and Computing, Tianjin University}
  \city{Tianjin}
  \country{China}
}
\email{senchen@tju.edu.cn}

\author{Lingling Fan}
\affiliation{%
  \institution{College of Cyber Science, Nankai University}
    \city{Tianjin}
  \country{China}
}
\email{linglingfan@nankai.edu.cn}

\author{Chunyang Chen}
\affiliation{%
  \institution{Monash University}
  %\city{Australia}
  \country{Australia}
}
\email{chunyang.chen@monash.edu}

\author{Xiaohong Li}
\affiliation{%
  \institution{College of Intelligence and Computing, Tianjin University}
  \city{Tianjin}
  \country{China}
}

\begin{abstract}
Approximately 15\% of the world's population is suffering from various disabilities or impairments. However, many mobile UX designers and developers disregard the significance of accessibility for those with disabilities when developing apps. It is unbelievable that one in seven people might not have the same level of access that other users have, which actually violates many legal and regulatory standards. On the contrary, if the apps are developed with accessibility in mind, it will drastically improve the user experience for all users as well as maximize revenue. Thus, a large number of studies and some effective tools for detecting accessibility issues have been conducted and proposed to mitigate such a severe problem.

However, compared with detection, the repair work is obviously falling behind. Especially for the color-related accessibility issues, which is one of the top issues in apps with a greatly negative impact on vision and user experience. Apps with such issues are difficult to use for people with low vision and the elderly. Unfortunately, such an issue type cannot be directly fixed by existing repair techniques. To this end, we propose \tool, an automated {and context-aware} repair method to fix the color-related accessibility issues (i.e., the text contrast issues and the image contrast issues) for apps. By leveraging a novel context-aware technique that resolves the optimal colors and a vital phase of attribute-to-repair localization, \tool not only repairs the color contrast issues but also guarantees the consistency of the design style between the original UI page and repaired UI page. Our experiments unveiled that \tool can achieve a 91.38\% repair success rate with high effectiveness and efficiency. The usefulness of \tool has also been evaluated by a user study with a high satisfaction rate as well as developers' positive feedback. 9 of 40 submitted pull requests on GitHub repositories have been accepted and merged into the projects by app developers, and another 4 developers are actively discussing with us for further repair. \tool is publicly available to facilitate this new research direction.
\end{abstract}




\maketitle

\section{Introduction}\label{sec:intro}
Nowadays, mobile applications (apps) are ubiquitous~\cite{chen2019storydroid,chen2022storydistiller,zhang2023web,zhang2023scenedroid}. In addition to providing various functional services for users, the importance of mobile accessibility has gained increasing attention from both industry and academia~\cite{importance,vendome2019can,important20,important21}. Mobile accessibility refers to making apps more accessible to people with disabilities when they are using mobile phones~\cite{definition}. Besides its special significance for the disabled, if the developers design a mobile app that has more {accessible features} such as screen readers
(TalkBack for Android~\cite{talkback}, VoiceOver for iOS~\cite{voiceover}), they will be able to reach a larger number of audiences. 
To this end, governments have established laws to help eliminate barriers in electronic and information technology for people with disabilities~\cite{web:USlaw, web:EuropeLaw} and leading IT companies (e.g, Google, Apple, Microsoft, Meta, and IBM) have established their accessibility teams to improve app accessibility~\cite{GoogleAccessibility,AppleAccessibility,IBMAccessibility,MicrosoftAccessibility,FacebookAccessibility}.

In recent years, a large number of empirical studies have been conducted to investigate the characteristics of app accessibility~\cite{ross2018examining,vendome2019can,yan2019current,ross2020epidemiology,alshayban2020accessibility,chen2021accessible,naranjo2022preliminary,da2022accessibility}. These studies unveiled that almost all apps are suffering from accessibility issues~\cite{alshayban2020accessibility,chen2021accessible,lopes2022can}. To mitigate such a severe problem, a series of effective automated approaches for detecting app accessibility issues have been proposed such as {Android Lint~\cite{lint}, Espresso~\cite{web:androidEspresso}, Robolectric~\cite{web:androidRobolectric}, Google Accessibility Scanner~\cite{scanner}, Google Accessibility Testing Framework (ATF)~\cite{framework}, MATE~\cite{eler2018automated}, LabelDroid~\cite{chen2020unblind}, AccessiText~\cite{accessiText}, Latter~\cite{salehnamadi2021latte}, Groundhog~\cite{salehnamadi2022groundhog}, and {Xbot}~\cite{chen2021accessible}}. Yet, too many accessibility issues make it difficult to effectively repair in practice. For example, the result in~\cite{chen2021accessible} shows that there are 40 issues for each app and 6.5 issues for each page on average. In addition to the bottlenecks from a large number of issues, the various categories of issues further limit repair efficiency. In other words, it is unrealistic for app developers to repair these issues within apps even with substantial human effort (also proved by the developers' feedback in~\S~\ref{subsec:rq3}).

\begin{figure}
\centering
\includegraphics[width=0.45\textwidth]{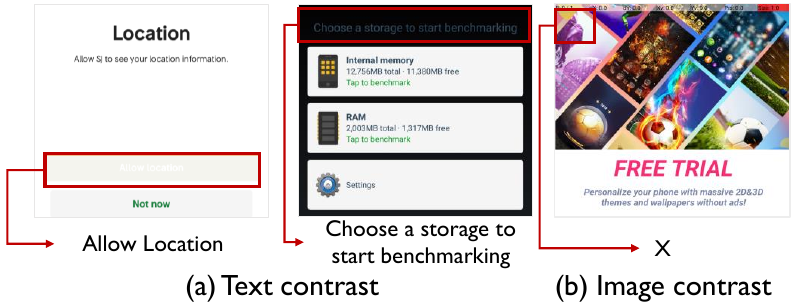}
\caption{Examples of color-related accessibility issues in apps.}
\label{fig:examples}
\end{figure}

To address this problem, researchers tried to fix these issues by leveraging automated repair approaches, but such effort is in its infancy.
Specifically, Alotaibi et al.~\cite{alotaibi2021automated} proposed a genetic algorithm guided by a fitness function to automatically repair size-based accessibility issues in apps. 
Moreover, the issues related to item labels (i.e., missing content labels) can be mitigated by different existing approaches such as social annotation techniques~\cite{zhang2018robust}, deep learning algorithms~\cite{chen2020unblind,mehralian2021data}, and crowd-sourcing techniques~\cite{brady2015crowdsourcing}. 
However, in addition to the above two issue categories with a large proportion, the issues of text or image contrast are also very serious, which is one of the most prevalent accessibility issues that affect mobile apps~\cite{ross2020epidemiology,alshayban2020accessibility,chen2021accessible,colorContrastImportant}. As shown in~\Cref{fig:examples}, text or image contrast, also known as {\textit{color-related accessibility issues}}, occurs when the color contrast between the text/image and the background is less than the minimum ratio specified by the accessibility guidelines~\cite{WCAG,colorContrastImportant}. Such issues make the apps difficult to use, not only for people with low vision but also for all users. 
{Linares-Vásquez et al.~\cite{linares2018multi} proposed a method to generate a brand-new color scheme for the UI, intending to reduce the energy consumption of the GUI in Android apps. Although it also changes the color of the UI page, it is completely different from its original color scheme, and the scenario of their work is completely different from this paper.}

{There are also techniques for repairing accessibility issues of web pages~\cite{mahajan2018automated,mahajan2018automated2,mahajan2017xfix,mahajan2017automated,panchekha2016automated}. However, they focused more on Mobile Friendly Problems, such as Font sizing, Tap target spacing, Content sizing, Viewport configuration (i.e., sizing issue), and Flash usage (i.e., rendering issue), which can inspire the repair of size-related issues, however, cannot benefit color-based accessibility issues.}
{Moreover, to help meet contrast requirements on web pages, several works~\cite{sandnes2021inverse,hansen2019still} focused on recommending color pairs by simply adjusting the color values of texts, however, they did not pay attention to the original design style of the web page.}
{Additionally, the implementation mechanisms are significantly different for web apps and Android apps, which directly distinguish repair solutions.}

Thus, it is actually a non-trivial task due to the following challenges: \textbf{C1:} The color-related changes should ensure to maintain the consistency of the design style between the original UI pages and repaired UI pages. \textbf{C2:} The UI components with issues and their corresponding attribute to be modified need to be accurately located and determined. Meanwhile, it should be determined whether the involved image files that are relevant to the image contrast issues really need to be modified. \textbf{C3:} The repaired results need to be evaluated and confirmed by real users and developers.

To address these above challenges, we proposed {\tool}, an automated {and context-aware} approach to repa\underline{\textbf{i}}ring the colo\underline{\textbf{r}}-related access\underline{\textbf{i}}bility i\underline{\textbf{s}}sues for Android apps. Specifically, to address \textbf{C1}, we design a novel 
{context-aware technique that resolves the optimal colors} by leveraging the well-designed criteria for color value resolving and the color reference DB construction. The 
{context-aware} technique can ensure the design style consistency between the repaired UI pages and original UI pages for apps. For \textbf{C2}, taking the selected colors as input, we use {an attribute-to-repair localization method} by analyzing the source code of relevant layout files to determine the component attributes that should be modified.
{Meanwhile, the repair constraints parsing step helps to determine if the image contrast issues need to be really fixed.} 
Based on these key phases, the issues within the apps can be effectively repaired and validated before a new repaired APK (Android Application Package) file is released. Last but not least, to address \textbf{C3}, we carry out a comprehensive and well-designed user study to help evaluate our repaired results. We also submit a number of pull requests for real GitHub projects to help improve the accessibility of their apps. 

To evaluate the effectiveness, efficiency, and usefulness of \tool, we designed a series of experiments {on 100 real-world apps, including both closed-source and open-source apps}. The results show that \tool performs a high repair success rate of 91.38\% and costs 2.27 minutes per app on average. Finally, based on the user study results and the developers' feedback, we highlight that \tool can really help developers fix color-related accessibility issues and practically improve the app accessibility. The user study results also show that the consistency of the original design style is well maintained from the perspective of both the UI page and the app level. {Till now, we have submitted {40} pull requests on GitHub repositories and 9 projects have merged our repaired results, and another 4 developers are actively discussing with us for further repair.}

In summary, we make the main contributions as follows.

\begin{itemize}
    \item {To the best of our knowledge, \tool is the first automated approach proposed to repair color-related accessibility issues for Android apps.} 
    We make the tool and relevant data public available on GitHub,\footnote{\url{https://github.com/iris-mobile-accessibility-repair/iris-mobile}} to facilitate new research areas for improving app accessibility.
    
    \item We propose a novel context-aware technique that resolves the optimal colors to ensure the consistency of design style between the required UI pages and the original UI pages.
    Additionally, a color reference DB collected from 9,978 apps has been constructed and released to help resolve the optimal color value for the color selection.
    
    \item The experiments on 100 real-world apps including both closed-source apps and open-source apps demonstrate the effectiveness and efficiency of our approach. A well-designed user study clearly demonstrates the usefulness of our approach. Moreover, the positive feedback from real developers has also highlighted the practicality of \tool.
\end{itemize}

\section{Preliminary}
\subsection{\textbf{Color-related Accessibility Issue}}
As shown in \Cref{fig:examples}, color-related accessibility issue includes two types: \textit{text contrast issue} and \textit{image contrast issue}. The former corresponds to visible text, where there is a low contrast ratio between the text color and background color. The latter refers to images with a low contrast ratio between the foreground and background colors. The visual presentation of text and text images has a contrast of at least 4.5:1 {(text below 18 point regular or 14 point bold)}, while the contrast of large-scale text {(18 point and above regular or 14 point and above bold)} and large-scale text image is at least 3:1 (required by Web Content Accessibility Guidelines (WCAG)~\cite{WCAG} and Google Accessibility Guidelines for Android~\cite{android_access_guideline}).
These two types of issues frequently occur in apps~\cite{alshayban2020accessibility,chen2021accessible,colorContrastImportant} and significantly reduce the app accessibility. Specifically, the examples shown in \Cref{fig:examples} have a big visual problem even for all people, not just for people with disabilities (e.g., people with low vision and the elderly).

\begin{figure}
\centering
\includegraphics[width=0.325\textwidth]{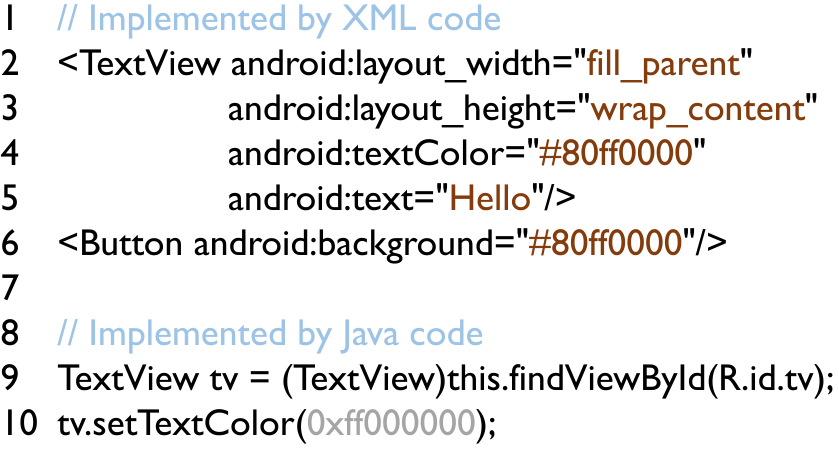}
\caption{Examples of color-related layout implementation.}
\label{listing:implementation}
\end{figure}

\subsection{\textbf{Color-related Layout Implementation}}
There are 2 main ways to implement the color setting of UI components~\cite{document}. As shown in \Cref{listing:implementation}, we can use XML layout code to set the color property for different UI components. For example, we can use ``android:textColor'' (Line 4) to draw the color for the text of \textit{TextView} and ``android:background'' (Line 6) to config the color for the background of \textit{Button}. The same functionalities can be completely replaced by Java code by using the corresponding APIs like \textit{\#setTextColor()} (Line 10).

\begin{figure}
\centering
\includegraphics[width=0.45\textwidth]{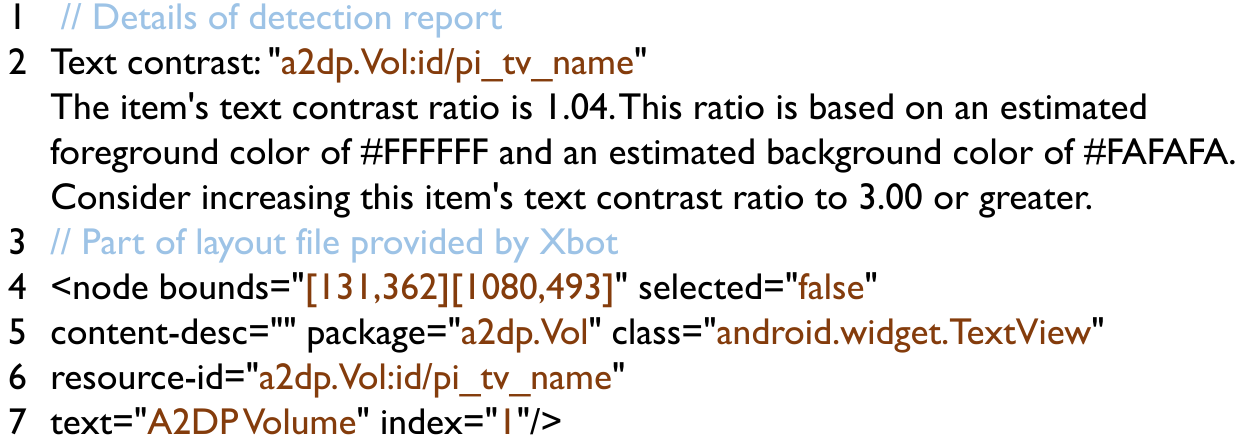}
\caption{The example of report generated by {Xbot}~\cite{chen2021accessible}.}
\label{listing:report}
\end{figure}

\subsection{\textbf{Reports Collected from Detection Tools}}
{The input required for repair needs to be obtained {from} the Google official accessibility testing framework (ATF)~\cite{framework}, a library collecting variable accessibility-related checks on View objects as well as AccessibilityNodeInfo objects.
Among the existing accessibility issue detection tools using ATF~\cite{Google-Monkey,scanner,eler2018automated,web:androidRobolectric,web:androidEspresso,accessiText,salehnamadi2021latte,salehnamadi2022groundhog}, {Xbot}~\cite{chen2021accessible}, which is a fully automated approach for detecting all types of accessibility issues based on the ATF and Google Accessibility Scanner~\cite{scanner}, has the ability to explore the app UI with high activity coverage, and can effectively and efficiently collect a relatively comprehensive dataset of accessibility issues. 
{Therefore, we finally choose {Xbot} as our {issue detection and collection} tool, which takes as input an APK file and outputs its exploration and detection results.}
\Cref{listing:report} is a partial example of the detection report and layout file provided by {Xbot}. The detection report prompts the type of the accessibility issue (i.e., text or image contrast), the unique identification of the component (i.e.,~{\textit{resource-id}} and {\textit{node bounds}}), and the specific information of the issue. Moreover, {Xbot} also provides the rendered UI screenshot for each activity.
}

\subsection{\textbf{Default Solution in Android}}
Android OS provides support for addressing color-related accessibility issues~\cite{android.highcontrast}. The repair strategy is that the system setting of ``High Contrast Text'' will change the UI components of used apps to black or white according to the original color, trying to make the text on the device easier for the user to read. However, after this setting is turned on, the text of all colors in the app will be changed into white or black, which will completely change the design style of the UI pages.

\section{Approach}
\begin{figure}
\centering
\includegraphics[width=0.5\textwidth]{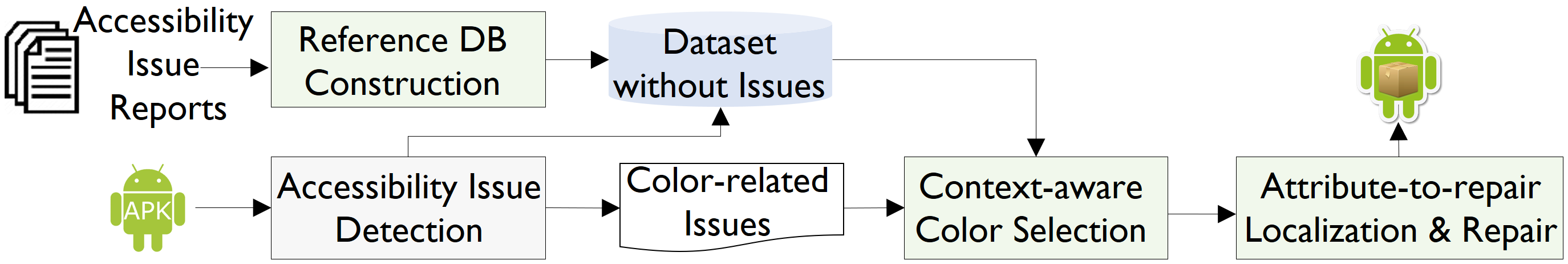}
\caption{Overview of \tool.}
\label{fig:overview}
\end{figure}

\Cref{fig:overview} shows an overview of \tool, {which takes as input an APK file along with the issue reports}, and outputs a repaired APK file without color-related accessibility issues. 
There are no special restrictions on the input APK, while the required reports are provided by ATF~\cite{framework}.
\tool consists of three main phases: \textit{\textbf{(1)}} \textit{Reference DB construction}, which analyzes the UI components without color-related issues from a large number of detection results, and constructs a reference database to further help select optimal color. \textit{\textbf{(2)}} \textit{{Context-aware} color selection}, {which is a novel context-aware technique that resolves the optimal value of the color replacement through two strategies based on our well-designed criteria for color value resolving.}
\textit{\textbf{(3)}} \textit{Attribute-to-repair localization}, which is used to locate the position of relevant UI components and further determine the attributes of components that need to be modified. After that, \tool replaces the attributes of the problematic UI component with the resolved optimal color value and updates the corresponding layout files or source code to repair the app. 

\subsection{\textbf{Reference DB Construction}}\label{subsec:db}
To provide reference values for selecting the optimal value of the replacement colors, the goal of this phase is to {construct a reference database containing a dataset without color-related issues (i.e., meet the requirements of standard color contrast). \textit{We highlight the replacement color selected from the database that has been used in real apps and accepted by the app designers.}}

Based on our investigation of the detection reports, we notice that if we only consider the color used by the app itself, it may be difficult to find a replacement color that can be applied, since most UI components in an app generally share the same color value. 
Therefore, it is essential to construct a database composed of multiple APKs instead of only relying on an app itself. 
Meanwhile, we investigated that different types of UI components (e.g., \textit{Button, EditText,} and \textit{TextView}) have different display styles. Therefore, the influence of the category of components needs also be thought about in the reference database construction. 
Based on these primary investigations, \textit{\textbf{(1)}} Owing to a significant number of accessibility issue reports detected by Xbot~\cite{chen2021accessible}, we first collect all UI components without contrast issues for all selected APKs. 
Meanwhile, we categorize this dataset by the UI component type. 
\textit{\textbf{(2)}} In a similar manner, {for each APK, we also collect the UI components without issues.}

The purpose of building the database is to obtain the color pairs composed of the foreground color and background color of each UI component. For the component dataset without issues, the reports do not involve the specific value of the color pair of UI components, so we need to make further efforts to compute the value of the color pairs. In this process, we extract the values of the key foreground and background colors from the screenshot of each UI component.
{The color composition of a single component is relatively simple and usually consists of only the background color and the color of the text or image on it.}
Therefore, we use \textit{\#getcolors()}~\cite{getcolors} in the image module to return the two most used colors in the screenshot. After judging that the result meets the required contrast, the color pair is returned as a replacement value for reference.

\subsection{\textbf{Context-aware Color Selection}}\label{subsec:color}
{Given a component with color-related accessibility issues,}
the goal of this phase is to determine the optimal replacement color for components, which do not violate the standard color contrast on the premise of maintaining the style of the original design as much as possible {(i.e., context-aware)}.
\textit{This is also the innovation of our approach, that is, to ensure style consistency at all design levels.}
When resolving the optimal replacement color, two aspects need to be considered: \textit{\textbf{(1)}} For a component with low contrast, how to decide whether to change the foreground color or background color, or both? \textit{\textbf{(2)}} 
What is the strategy of color selection and 
what are the criteria for color value resolving under the selection strategy?

For question \textit{\textbf{(1)}}, many UI components usually share an area with the same background color or call the same resource defining the background color, on one page of an app. Therefore, choosing to modify the value of the background color is too easy to have a chain reaction, which may change the color contrast between the foreground and background of other UI components, or even lower than the standard value, and introduce new accessibility issues. The risk caused by this modification is high. Compared with the background color, the foreground color is relatively independent, usually expressed as the color of the text in the component or the main color of a picture. When the foreground color changes, it would probably not trigger changes in other parts of the UI page. Therefore, our approach takes priority to modifying the foreground color of the problematic component with the goal of obtaining the optimal solution for the replacement color. 

\begin{figure}
\centering
\includegraphics[width=0.35\textwidth]{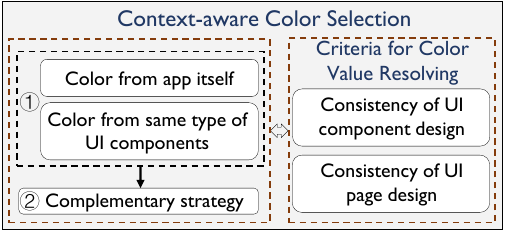}
\caption{The workflow of context-aware color selection.}
\label{fig:Adaptive-Color-Selection}
\end{figure}

After determining the target to be modified, the next problem to be solved is to find the replacement color required for modification. The most important thing is to keep the original UI style unchanged. Therefore, for question \textit{\textbf{(2)}}, as shown in \Cref{fig:Adaptive-Color-Selection}, we propose two strategies of color selection.

\subsubsection{Color Selection Strategy} Two situations:
\textbf{\ding{172}} the color replacement based on the reference DB. Through this strategy, we can get the replacement color from the app itself or the dataset filtered by the same component types in other apps. Specifically, if we can find alternative colors from the app itself, we can directly use the colors defined and used by this app itself for replacement, which is more in line with the color selection intention of the app designers. If no suitable replacement color can be found in the app itself, we will consider the same component types collected from other apps. Because the same types of components probably share similar design styles. \textbf{\ding{173}} Complementary strategy will be applied if we cannot find a replacement color from the reference DB. This strategy chooses the replacement color by directly modifying the original foreground color for the problematic component while maintaining the design style of the UI component and the coordination of the UI page.

\subsubsection{Criteria for Color Value Resolving (Context-aware)}
Note that the replacement color matched from the reference DB by using the background color of the problematic component is a set of color candidates. Meanwhile, a reference is also required to directly modify the value of the original color when using the complementary strategy. Thus, 
%\revise{to maintain the original design style of the app as much as possible, we define the color value resolving as a kind of context-aware technique, which includes two criteria.}
{to maintain the original design style of the app as much as possible, we define two criteria for color value resolving.}
\ding{172} Consistency of UI component design: the hue and saturation level are consistent with the original color of UI component. \ding{173} Consistency of UI page design: the hue is more harmonious with the overall hue of the UI page.

\noindent{\textbf{Consistency of UI component design.}}
{From the perspective of component design, the color matching of components represents the designer's intention.
UI designers usually first consider the hue of components when setting colors for them, such as the color of red or blue. In addition, colors with the same hue but different saturations also have great visual differences. 
{Actually, various saturations produce a variety of visual impacts and attractions. For example, a color with high saturation is bright, which can make the main body stand out from the background, while a low saturation color can give people a low-key and subtle feeling.} 
Therefore, when considering the designer's intention to the greatest extent and ensuring that the replacement color is relatively consistent with the original color, it is crucial that the hue and saturation level of the replacement color and the original color should be consistent.}
We use HSV (hue-saturation-value)~\cite{ong2014comparative} to meet such a criterion since the HSV color model is consistent with the way humans describe colors, and allows independent control of hue, saturation, and intensity (value). The HSV values of colors can be calculated using their RGB values by formulas (\ref{eq1})$\sim$(\ref{eq3})~\cite{smith1978color}. 
\begin{equation}
\label{eq1}
H= \left \{
\begin{array}{lr}
    0, \ Max = Min\\
    60 \times \frac{G - B}{Max - Min}, \ Max = R\\
    60 \times \frac{B - R}{Max - Min} + 120, \ Max = G\\
    60 \times \frac{R - G}{Max - Min} + 240, \ Max = B
\end{array}
\right.
\end{equation}
\begin{equation}
\label{eq2}
S= \left \{
\begin{array}{lr}
    0, \ Max = 0\\
    \frac{Max - Min}{Max}, \ Max \neq 0
\end{array}
\right.
\end{equation}
\begin{equation}
\label{eq3}
V = Max 
\end{equation} 
where R, G, and B represent red, green, and blue values of the RGB of one color, and ``Max'' and ``Min'' represent the maximum and minimum values between R, G, and B values. These formulas operate on values in the form of decimal numbers.
Based on this standard, according to the Practical color coordinate system (PCCS)~\cite{saito1996comparative,guan2002study}, we divide the calculated saturation value from 0 to 1 into three saturation levels: low (0$\sim$0.33), medium (0.34$\sim$0.67), and high (0.68$\sim$1). When filtering the replacement color values, we only retain the candidate values with the same hue and saturation level as the original color.

\begin{figure}
\centering
\includegraphics[width=0.4\textwidth]{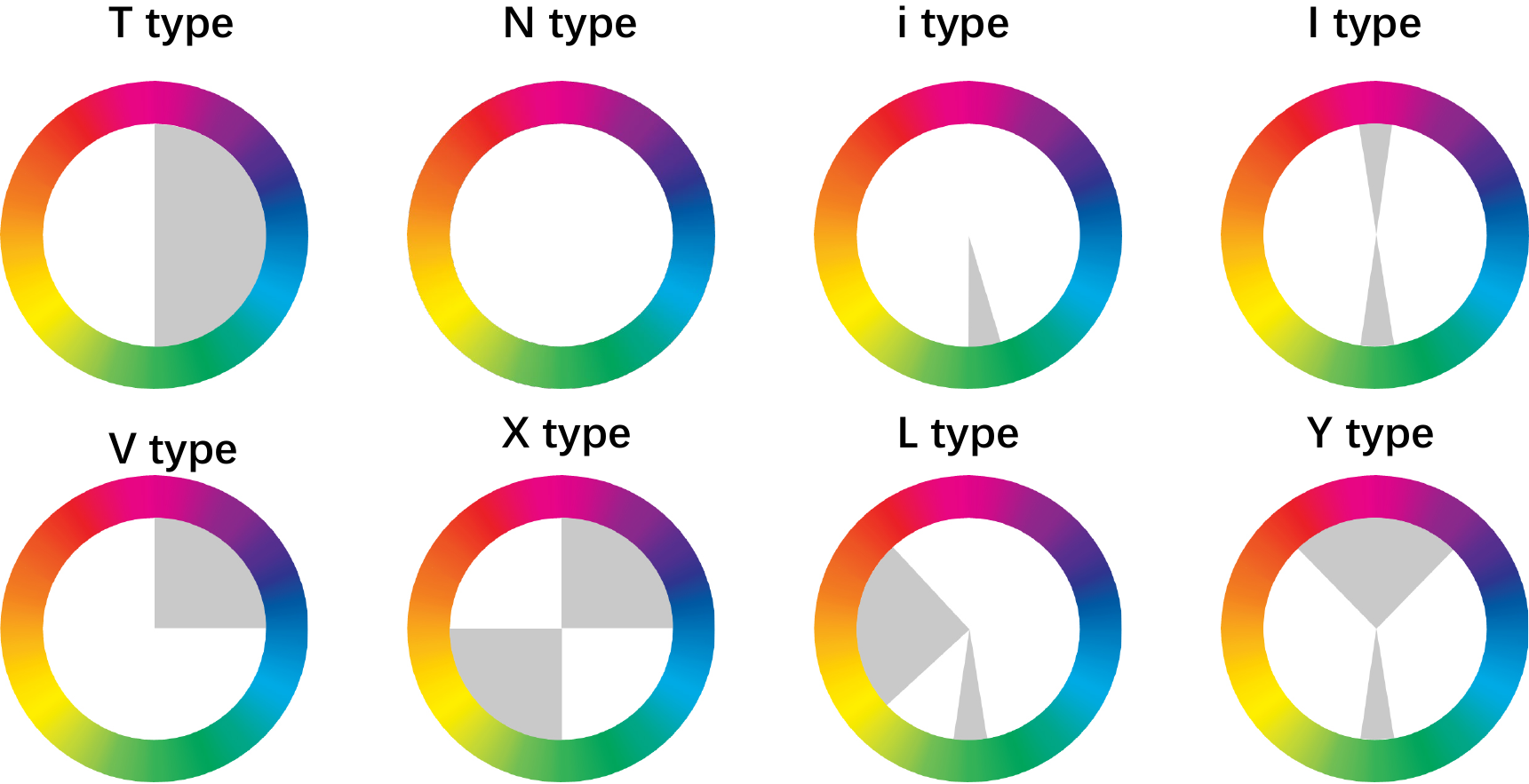}
\caption{Eight harmonic types for color value resolving.}
	\label{fig:harmonic_types}
\end{figure}

%\smallskip
\noindent{\textbf{Consistency of UI page design.}} As shown in \Cref{fig:harmonic_types}, when considering the color coordination degree of the whole UI page, we leverage eight harmonic types defined on the hue channel of the HSV color wheel as the second criterion~\cite{cohen2006color}. Each type shows the hue color distribution in the harmonic template (the size of the gray area is fixed, but the position is not fixed, and it can rotate around the center of the circle). In other words, if all the hues of a UI page fall in the gray area of a certain harmonic type, the color replacement of the page is considered to be harmonic.

{Using} the two strategies with two criteria defined in \Cref{fig:Adaptive-Color-Selection}, we can get the final optimal color for replacement. 
As shown in Algorithm~\ref{algo:color}, before calculating the optimal replacement color value, \tool will obtain the set of colors available for replacement from the reference DB ($ColorSet_{ref}$) and calculate the optimal harmonic type and deflection angle corresponding to the UI page ($best_{T}$ and $best_{Alpha}$). After obtaining the inputs, we calculate the HSV value of the original color first (Line 1). Then, by judging whether the hue and saturation level of the replaceable candidate color is consistent with the original color {(\textit{Consistency of UI component design})}, a \textit{ColorSet} (Line 2$\sim$Line 5) that meets the consistent color hue and saturation standard is selected from the set of replaceable candidate colors. If the \textit{ColorSet} is not empty, the distance between each color in the \textit{ColorSet} and the shadow part of the optimal harmonic type is calculated (Line 8). The smaller the distance is, the more consistent the hue consistency standard of the UI page is \textit{(Consistency of UI page design)}. The color with the smallest distance is selected from the \textit{ColorSet} as the optimal replacement color (Line 10).

If the \textit{ColorSet} is empty, {it means no appropriate color can be selected for replacement from the existing reference DB.}
To find a suitable replacement color, 
we use the complementary strategy to directly modify the HSV value of the original color.
If the original color is black, white, gray, and other achromatic system colors (neutral colors), since there is only a difference in brightness between them, it is only necessary to adjust the brightness (V) value of the original color up and down until the changed color meets the standard color contrast (Line 13).
At this time, since the changed color is still neutral, the hue (H) channel of the HSV color wheel of the whole UI page will not be affected \textit{(Consistency of UI page design)}.
The saturation (S) value of the original color is also not changed. But if the original color belongs to the color system (not neutral colors), adjust the value of H and S on the premise that the hue and saturation level is consistent with the original color \textit{(Consistency of UI component design)}:  
(1) Adjust the value of H first, and then adjust the value of S to obtain the replaceable value $color_{HS}$ (Line 15). (2) Adjust the value of S first, and then adjust the value of H to obtain $color_{SH}$ (Line 16). Finally, the color closest to the original color is selected as the optimal replacement color among the two replaceable colors (Line 17).

\begin{algorithm2e}[t]\footnotesize
\setcounter{AlgoLine}{0}
\caption{Context-aware Color Selection Algorithm}
\label{algo:color}
\DontPrintSemicolon
\SetCommentSty{mycommfont}
\KwIn{$Color_{org}$: The problematic color with contrast issue\;
$ColorSet_{ref}$: The color candidates selected from reference DB\;
$best_{T}$: The optimal harmonic type of the corresponding UI page \; 
$best_{Alpha}$: The deflection angle of the optimal harmonic type \;
\KwOut{$Color_{opt}$: The selected optimal color\;}
$H_{0}$,$S_{0}$,$V_{0}$ $\leftarrow$ $getHSV$($Color_{org}$) \;
\ForEach{$c$ $\in$ $ColorSet_{ref}$}{
    $H$,$S$,$V$ $\leftarrow$ $getHSV$($c$) \;
    \If{$isConsistentHue$($H_{0}$, $H$) and $isConsistentSaturation$($S_{0}$, $S$)}{
        $ColorSet$.append($c$) \;
        }
	}
\If{$ColorSet$ is not null}{
	\ForEach{$c$ $\in$ $ColorSet$}{
	    $d$ $\leftarrow$ $getDistance$($best_{T}$, $best_{Alpha}$, $c$) \;
	    $DisSet[c]$ = $d$ \;
	    }
	 $Color_{opt}$ $\gets$ $minDistance$($DisSet$) \;
	 }
\Else{
    \If{$isNeutralColor$($Color_{org}$)}{
	    $Color_{opt}$ $\gets$ $adjustV$($V_{0}$,$Color_{org}$) \;
	    }
	 \Else{
	     $Color_{HS}$ $\gets$ $adjustHS$($H_{0}$, $S_{0}$, $Color_{org}$) \;
	     $Color_{SH}$ $\gets$ $adjustSH$($H_{0}$, $S_{0}$, $Color_{org}$) \;
	     $Color_{opt}$ $\gets$ $minChanged$($Color_{org}$, $Color_{HS}$, $Color_{SH}$) \;
	    }
    }

\Return $Color_{opt}$\;
}
\end{algorithm2e}

\subsection{\textbf{Attribute-to-repair Localization}}\label{subsec:localization}
{After obtaining the optimal color for repairing UI components, this phase aims to localize the components and determine the attributes to be repaired.} 
Although the input detection report contains specific information about accessibility issues, the specific location and the attributes that need to be modified cannot be directly determined. 
There are several challenges. \textit{\textbf{(1)}} First of all, color-related accessibility issues include two types of problems. 
These two types are different in repair objects and repair methods. For example, for text contrast issues, we need to repair the color attribute of the text in the UI components. For the issues of image contrast, we need to replace or modify the involved images. \textit{\textbf{(2)}} Secondly, even for the same type of issues, the attributes and the repair conditions to be modified will be different. \textit{\textbf{(3)}} In addition, how to decide the location of the layout code (or source code) of the relevant components in the UI layout files of the app through the existing detection report is also a big challenge.
{In this phase, \tool treats these two types of issues separately and decides the attributes to be modified by analyzing both the layout code and the source code.}

\subsubsection{\textbf{Localization of the related UI components}}
{The input report contains two types of tips about the information of the components: ``Component ID'' and ``Bounds'' of a component in the layout. Note that the layout file here refers to the layout file gained by Xbot, denoted by \textit{Xbot-Layout}, which is different from those obtained by decompiling the APK file, denoted by \textit{Decompile-Layout}. To localize the relevant components to be repaired, we analyze these two different types of information. \textit{\textbf{(1)}} For component ID, since ID is unique to objects, 
we can directly locate the property set of the corresponding component in the \textit{Decompile-Layout} (or source code) according to the component ID (\Cref{fig:attribute-img}). \textit{\textbf{(2)}} For the bounds of a component, the report displays Bounds instead of ID because 
Xbot sometimes failed to detect the component ID. Meanwhile, the attribute ``bounds'' does not exist in the \textit{Decompile-Layout} files. Thus, to locate the relevant components, we can find the text information of the component according to the bounds in \textit {Xbot-Layout}, and choose to match through the \textit{Android:text} attribute in \textit{Decompile-Layout} (the steps 1$\sim$3 in \Cref{fig:attribute-text}).}

\subsubsection{\textbf{Acquisition of the related attribute-to-repair}}\label{subsub:acq}
\begin{figure}
\centering
\includegraphics[width=0.5\textwidth]{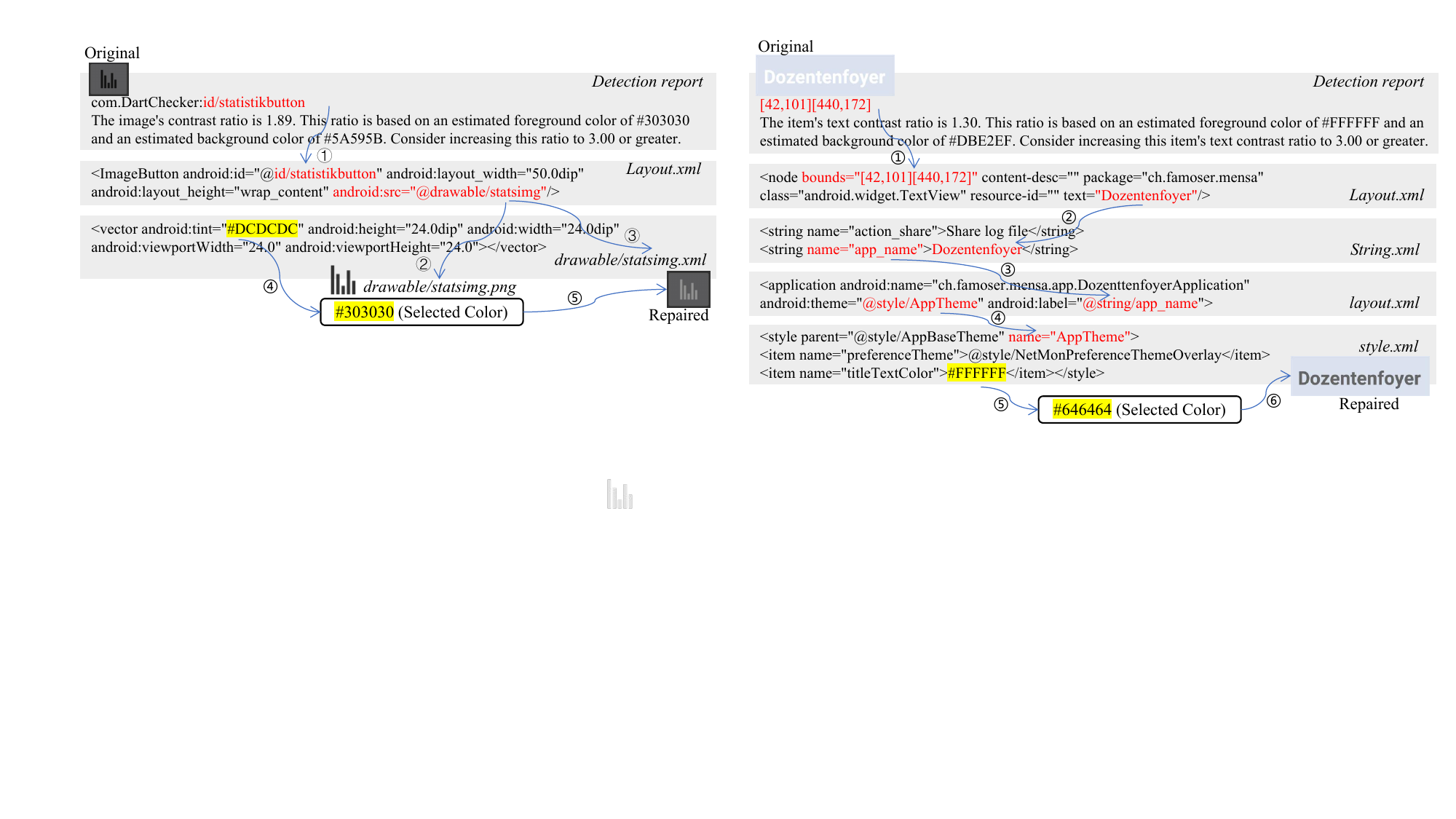}
\caption{Example of attribute-to-repair localization for text contrast.}
\label{fig:attribute-text}
\end{figure}

\begin{figure}
\centering
\includegraphics[width=0.5\textwidth]{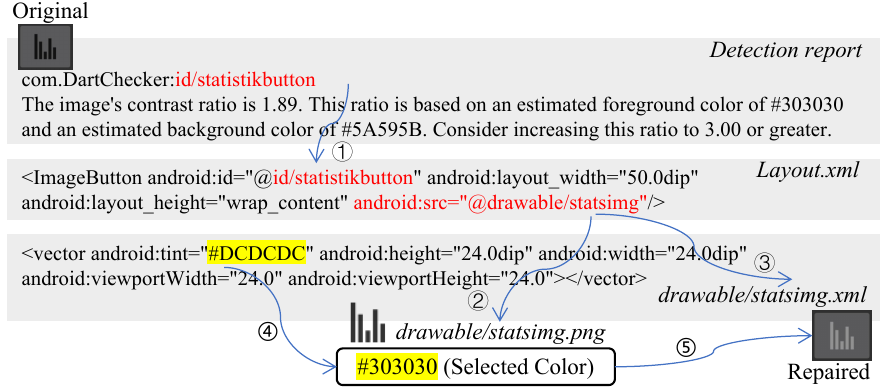}
\caption{Example of attribute-to-repair localization for image contrast.
% \revise{@yuxin, change to pdf}
}
\label{fig:attribute-img}
\end{figure}

\tool uses static data-flow analysis to extract relevant attribute sets shown in~\Cref{fig:attribute-text} and \Cref{fig:attribute-img}. Firstly, 
{\tool parses the app to extract the layout files and \textit{Smali code}.}
Then, through the above analysis, \tool completes the localization of the relevant components and obtains the attribute sets for each component. For the two types of accessibility issues, the attributes-to-repair are different. \textit{\textbf{(1)}} For text contrast issues, the components are generally \textit{TextView}, \textit{EditText}, and \textit{Button}, which are prone to such problems. At this time, the foreground color refers to the color of the text in the component, and the background color is the color of the component background. Therefore, for such issues, the object to be repaired is mainly the color of the text in these components. However, since the text in \textit{EditText} is mainly used for input hints and the text color is generally light, the color contrast of the component does not have to meet the standard requirements, and such components need to be filtered out in the localization stage. To conclude, the main attributes involved include \textit{Android:textcolor, Android:textcolorlink, Android:titletextcolor}, and other attributes, or set the attributes in the style file accordingly (the step 5 in~\Cref{fig:attribute-text}).
\Cref{fig:attribute-text} shows an example of attribute-to-repair localization for text contrast. \textit{\textbf{(2)}} For image contrast issues, the frequently occurring components mainly include \textit{ImageButton} and \textit{ImageView}. At this time, the issue with low contrast is the contrast between the image color and the background color of the component. 
Therefore, to prevent the impact on other components in the UI, we choose the image in the component as the repair object. The first step is to get the image files that need to be repaired. {Among them, the source file of an image can be directly associated with the UI component through the attribute definition in the layout file.} For example, as shown in~\Cref{fig:attribute-img}, images can be associated by setting the property values of \textit{Android:src} or \textit{Android:background} related to the components. \textit{Android:src=``@drawable/statsimg''} can be mapped directly to the file name in the resource folder (\textit{res/drawable/statsimg.png}). \Cref{fig:attribute-img} shows an example of attribute-to-repair localization for image contrast. At this time, the goal of repair is to change the color setting of the image in the component.
{However, some images are actually not suitable for modification, so we need to filter out them and conduct repairing, which is elaborated as follows.}

\subsubsection{\textbf{Repair constraints of images}} 
To distinguish the images suitable to be repaired, we divide these images into ``\textit{functional images}'' and ``\textit{ornamental images}'' according to the display intention. 
{We choose to modify the images or icons that focus more on using functions, such as return, add, cancel, and share icons, or images representing these meanings, which we refer to as ``functional images''.}
For the ``ornamental images'', if we change the color of the image, the display function of the image may be different from before. Therefore, we only choose to repair the ``functional images''. 
\tool distinguishes these two types of images by a lightweight static analysis technique that determines whether the images are associated with event handler methods. The trigger behavior of a component is usually associated with the ``clickable'' attribute in the layout file. Therefore, we judge whether the corresponding image is classified as a ``functional image'' by checking the relevant attributes of the problematic component.
{Then we execute the optimal color selection algorithm and change the color composition of the original image by changing the value of the color model~\cite{ibraheem2012understanding} of the pixel of the image, such as RGB~\cite{susstrunk1999standard} and RGBA~\cite{urban2019redefining}, which applies to  vector images in resources and other types of {user-uploaded non-vector images} such as PNG or JPG.}

After obtaining the replacement color and locating the attribute to be modified, \tool will directly replace the original values that have issues, including the replacement of color values and image files. \tool then repackages the replaced file to get a new APK.
Moreover, during repair, we also consider many other aspects such as the judgment of foreground and background color, to solve the possible detection errors of foreground and background color in the detection reports. 
{Sometimes, the set of the foreground color and the background color is reversed in the detection process, such as when the detection report indicates the foreground color as \#298670 and the background color as \#EDF064, when in reality, they are exactly the opposite. Before repair, we capture screenshots of the UI pages and extract the color information from them to ascertain the true color composition of each component.}

\begin{figure*}
\centering
\includegraphics[width=1\textwidth]{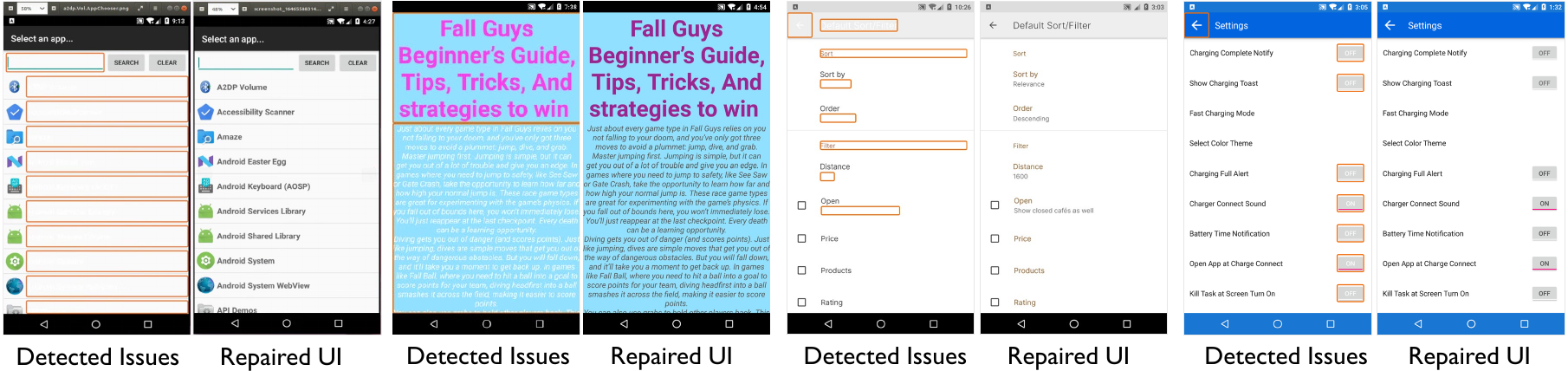}
\caption{Examples of repaired issues.}
\label{fig:repaired-examples}
\end{figure*}

\section{Experiments}
To make the experiments we designed better evaluate our approach, we raise the following questions: 

\begin{itemize}
    \item \textbf{RQ1:} How effective and efficient is \tool in repairing the color-related accessibility issues?
    \item \textbf{RQ2:} How much does each key strategy of \tool contribute to the overall performance?
    \item \textbf{RQ3:} How useful is \tool from the perspective of mobile users and app developers?
\end{itemize}

\noindent \textbf{Dataset.} 
For reference DB construction, we randomly collected 9,978 real apps, including 5,081 open-source apps from F-Droid~\cite{fdroid} and 4,897 closed-source apps from Google Play.
{For RQ1 and RQ2, considering the time cost of testing, we randomly selected 100 apps with an average size of approximately 6MB as experimental subjects, of which the number of open-source apps and closed-source apps is 50 and 50, respectively, and used \tool to automatically repair them.}
It should be noticed that these 100 apps do not appear in the dataset (i.e., 9,978) used in reference DB construction.

\subsection{\textbf{RQ1: Effectiveness and Efficiency Evaluation}}\label{subsec:rq1}
\subsubsection{\textbf{Setup}} To answer RQ1, we compare the detection results of 100 apps before and after repair. First, we use the Xbot~\cite{chen2021accessible} tool to conduct a preliminary detection on the tested apps and count the color-related accessibility issues of those apps before repair. Then, we take the detection results and APK files as input, use the \tool to repair the color-related accessibility issues of these apps one by one, repackage them, and output new APK files. To get the results after repair, we use the Xbot tool to detect the repaired APK files again and obtain the detection results. 
We evaluate the effectiveness of \tool by comparing the number of repaired issues and the issues in the same original app. 
The success rate for the $i^{th}$ app mentioned here is denoted by $RepairR$.

\begin{equation}
    RepairR_{i}=\frac{N_{i}^{Repaird_{issue}}}{N_{i}^{All_{issue}}}\times 100\%
\end{equation}

Additionally, to evaluate the efficiency of \tool, we record the execution time for these 100 apps and compute the average time to demonstrate the performance.

\begin{table}[t]
\centering
\small
\caption{Results for effectiveness evaluation of \tool.}
\scalebox{1}{\begin{tabular}{c|c|c|c}
\hline
\textbf{Issue Type} & \textbf{\begin{tabular}[c]{@{}c@{}}\# Real\\ Issues\end{tabular}} & \textbf{\begin{tabular}[c]{@{}c@{}}\# Repaired\\ Issues\end{tabular}} & \textbf{\begin{tabular}[c]{@{}c@{}}Success \\ Rate (\%)\end{tabular}} \\ \hline
\textbf{Text Contrast} & 660 & 618 & 93.6 \\ \hline
\textbf{Image Contrast} & 71 & 50 & 70.4 \\ \hline
\rowcolor{gray!20}\textbf{Total} & 731 & 668 & 91.38 \\ \hline
\end{tabular}}
\label{tab:RQ1_results}
\end{table}

\subsubsection{\textbf{Result}} The results of the effectiveness evaluation (RQ1) are shown in~\Cref{tab:RQ1_results}.
The column ``\# Real Issues'' represents the number of issues of the text and image contrast contained in all 100 apps. Note that, we removed a part of issues when counting the number of real issues for repairing. The removing parts include \textit{\textbf{(1)}} the \textit{EditText} components as we mentioned in~\S~\ref{subsub:acq}, \textit{\textbf{(2)}} the false positive cases caused by wrong screenshots in the detection reports, and \textit{\textbf{(3)}} the causes altered by the system design style (the problem also introduced in~\cite{alshayban2020accessibility}). Finally, we have 660 real issues with text contrast and 71 real issues with image contrast,
{which owns the most issues among all the issue types, accounting for 30.17\%.}
Remarkably, because a component, uniquely marked by ID, may be applied to different pages or the same page multiple times, there may be two or three issues caused by the same component. At this time, the property only needs to be modified once, which belongs to components with the same ID. 
As for text contrast, Table~\ref{tab:RQ1_results} shows that the number of issues of 100 apps before the repair is 660. Among them, there are {618} issues that have been repaired successfully, accounting for a {93.6\%} success repair rate and including {413} different UI components. The repair rate unveils that our tool can effectively reduce the number of text contrast issues. The root causes of the remaining {42} unresolved issues are as follows. \textit{\textbf{(1)}} For some components, although we have located their locations and initial attribute sets during the repair process and have tried to repair them, the user's actions may cause the components to appear in different states during the running of the app, failing the repair. Because the color rendering is implemented in complex source code supported by new reconstructed API interfaces of third-party libraries, rather than the official APIs introduced in~\Cref{listing:implementation}. \textit{\textbf{(2)}} Errors are due to the shortcomings of using bounds to locate attributes. In the process of issue detection, the detection tool cannot obtain the ID of some components, so we use the \textit{text} information of the component according to the \textit{bounds} to achieve positioning, and the positioning effect is not as good as using ID.

For the issues of image contrast, \Cref{tab:RQ1_results} shows that there are {71} image issues before repair, and {50} problems were repaired by \tool, including {35} different components. Therefore, the success rate of image contrast is {70.4\%}. There are still {21} issues that have not been solved for the reason of the shortcomings of using bounds to locate. Finally, the overall success rate of the two types of issues is {91.38\%}, and the total number of issues repaired is {668}. \Cref{fig:repaired-examples} presents several examples of the repair results by \tool. More examples can be found on our website~\cite{ourwebsite}.

Apart from evaluating the effectiveness of \tool in the number of repairs, we also record the execution time of \tool in practice. When repairing an APK, \tool will first use Xbot to detect the accessibility issues in it, update the reference DB, and then start to automatically repair. Finally, the average time of issue detection and DB updating is 100.7 seconds, and the average time of repair is 136.2 seconds. Compared with manual repair, \tool can greatly shorten the time of attribute localization and provide a feasible reference value for color replacement. Thus, \tool has high time efficiency.

\begin{table}
\small
\centering
\caption{Evaluation of the key phrase ({context-aware} color selection).}
\scalebox{1}{\begin{tabular}{c|c|c}
\hline
\textbf{Category} & \begin{tabular}[c]{@{}c@{}}\textbf{\# Repaired issues}\\ \textbf{by reference DB}\end{tabular} & \begin{tabular}[c]{@{}c@{}}\textbf{\# Repaired issues}\\ \textbf{by direct modification}\end{tabular}\\
\hline
\textbf{Text Contrast}  & 569 & 49 \\ \hline
\textbf{Image Contrast} & 39 & 11 \\ \hline
\rowcolor{gray!20}\textbf{Total} & 608 & 60 \\
\hline
\end{tabular}}
\label{tab:RQ2}
\end{table}

\subsection{\textbf{RQ2: Ablation Study}}\label{subsec:rq2}
\subsubsection{\textbf{Setup}} 
In RQ2, we aim to evaluate the key phases (i.e., {context-aware} color selection and attribute-to-localization). For the localization, the accuracy of this phase is consistent with $RepairR$ (i.e., 91.38\%). For {context-aware} color selection, there are two strategies to obtain the optimal replacement color: \textit{\textbf{(1)}} selected from reference DB; \textit{\textbf{(2)}} direct modification based on the complementary strategy. 
We also use Xbot to detect the accessibility issues of APK before and after repair {and investigate the number of repaired issues by each strategy.}

\subsubsection{\textbf{Result}} As shown in Table~\ref{tab:RQ2}, for the issue of text contrast and image contrast, the proportions of the two strategies are 569:49 and 39:11, respectively. 
{We can see that 91.02\% of the issues in the  selected 100 apps can be repaired by the reference DB, indicating the reference color DB plays an important role in the phase of {context-aware} color selection.} 
{Since the color pairs in reference DB are all from real apps, their color matching is recognized and loved by the designers and users of those apps, which also proves the rationality of our replacement.}
{That is also the reason why we prefer to utilize the reference DB to repair issues as many as possible at first.}
{Meanwhile, there are also some issues (i.e., 60) that cannot be repaired by the reference DB, while they can still be fixed by the complementary strategy. Although the replacement colors obtained using the complementary strategy do not appear in the apps in the existing DB, they also meet the standards, including the color contrast and two designed criteria.}

\subsection{\textbf{RQ3: Usefulness Evaluation}}\label{subsec:rq3}
To evaluate the usefulness of \tool, on the one hand, it is necessary to conduct a user study on the repaired UI page and repaired app. Meanwhile, the consistency of the design style between the original UI pages and repaired UI pages as well as the design style of the app level also should be evaluated by the validation from real users.
On the other hand, the pull requests that are accepted by real app developers on GitHub will make our repair results more convincing in a real scenario. During this process, the feedback from real developers facilitates the iterative improvement of \tool.

\subsubsection{\textbf{User study from app users}}

\noindent \textbf{Dataset.} To compare the effects before and after repair {from the app users' perspective}, \textit{\textbf{(1)}} we randomly selected 12 pairs of UI pages from the repaired UIs. Each pair of pages consists of the original UI and the repaired UI. Then, we divided the 12 pairs of pages into two groups as a control experiment (for \textit{Task 1}). \textit{\textbf{(2)}} To facilitate the scoring of participants, we randomly selected other 7 pairs of pages (for \textit{Task 2}). In each pair, besides the original UI page and the repaired UI page, we also added the UI page with the circled issues.
\textit{\textbf{(3)}} To evaluate the overall coordination of the design style of the repaired apps, we randomly selected 3 apps according to the different levels of the UI page number, whose UI page numbers are 4, 8, and 14 (for \textit{Task 3}). Among them, the number of issues in each app is not less than 10.

\begin{figure}
\centering
\includegraphics[width=0.45\textwidth]{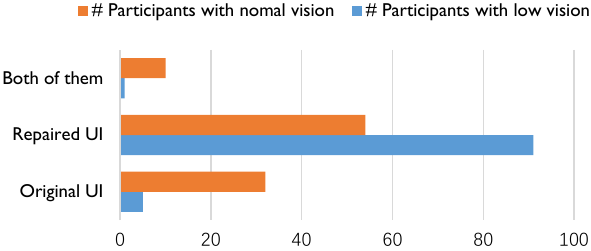}
\caption{Participants’ preference between the original UI pages and repaired UI pages.}
\label{fig:preference}
\end{figure}

\smallskip
\noindent \textbf{Participant recruitment.} 
We recruited 32 participants from our university, including 20 who are near-sighted (16 participants for \textit{Task 1} and \textit{Task 2} (both of them are near-sighted), and 16 participants for \textit{Task 3}).
None of them has used the repaired apps, which excludes the potential bias. The participants are from different countries, including Singapore, the United States, Germany, and China. These participants include undergraduates, postgraduates, PhD students, and staff.

\noindent {\textbf{Setup.}}  
{For app users, we design user research from two levels (i.e., the UI level and the app level). At the UI level, we investigate the user's preference for the UI pages before and after repair in \textit{Task 1} and the user's satisfaction with the repair effect of \tool in \textit{Task 2}.}
{The purpose of \textit{Task 1} is to demonstrate the usefulness of \tool for different users.}
To achieve it, we divided the page screenshots before and after repair into two groups, 
looked for people who are nearsighted with over 2.0 diopters as participants in this part,
and asked them to make preference choices for the two groups of screenshots without glasses (simulating people with special vision) and with glasses (normal vision). In this step, the order of each pair of screenshots is random and marked separately. 
{We request participants to replicate the real-world usage conditions as closely as possible when making their selections, which aims to accurately simulate the authentic experiences of diverse user groups.}

\begin{figure}
\centering
\includegraphics[width=0.45\textwidth]{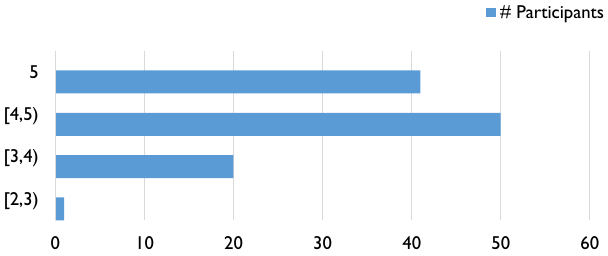}
\caption{Distribution of participants' scores.}
\label{fig:distribution}
\end{figure} 

The purpose of \textit{Task 2} is to test the repair effect of \tool through the score of participants. We showed the users the original UI page, the detection UI page with the issue identification (the problematic component is circled), and the repaired UI page. Meanwhile, we also explained to the participants what is the color-related accessibility issues and the goal of \tool. In this part, we randomly look for participants to rate the repair effect of \tool and explain the shortcomings of the repair or their suggestions. The main evaluation criteria are: 
\ding{172} Whether the issues are successfully repaired. \ding{173} Whether the design style of the UI page after the repair is consistent with the original design style. \ding{174} Whether the color matching of the whole UI page after the repair is coordinated. \ding{175} Whether the color changes are minimal. \ding{176} Score on the Likert scale ranging from 1 to 5. 

At the app level, we evaluate the overall coordination of the repaired app through user research (\textit{Task 3}). The goal of \textit{Task 3} is to verify whether \tool performs well in maintaining overall color harmony within the app. We asked the participants to circle the UI pages and uncoordinated components in each app and counted the proportion of repaired components among the circled components. It is worth noting that we set the same resolution and the same page size for each UI screenshot, ensure that each participant can see the same screenshot, and eliminate the impact of other factors on user selection.

\noindent {\textbf{User study result.}}
The results of the user study are shown in \Cref{fig:preference} and \Cref{fig:distribution}. \Cref{fig:preference} shows the preference of participants with different vision levels for the UIs before and after repair, in which the blue bar chart represents the preference of participants with low vision and the orange bar chart represents the choice of participants with normal vision. On the whole, our repair is more attractive to people with low vision. 
{Under the condition of low vision (without glasses), almost all participants chose the repaired page (94.80\%). Meanwhile, under normal vision, only one-third of the results of page {preferences} are the original pages, and more than half (56.3\%) of the results are the repaired pages.}
This also shows that the effect of our repair is also accepted and liked by most people, which can effectively solve the impact of low contrast on the viewing effect of people with weak vision. We also investigated the reasons why some participants chose the original page. For example, they said that they preferred the color matching of the original UI on the premise that they could see the text content. There is no doubt that this color preference has the participants' personal aesthetic standards.

\begin{table}[t]
\footnotesize
\centering
\caption{Results of \textit{Task 3}.}
\scalebox{1}{\begin{tabular}{ccccccc}
\hline
\multirow{2}{*}{\textbf{App Package}}& \multicolumn{3}{c}{\textbf{Component}} & \multicolumn{3}{c}{\textbf{UI Page}} \\
\cmidrule(lr){2-4} \cmidrule(lr){5-7}
& $C_{All}^{Re}$ & $C_{All}^{Un}$ & $C_{Re}^{Un}$ & $P_{All}^{Re}$ & $P_{All}^{Un}$ & $P_{Re}^{Un}$ \\
\hline
\hline
\textbf{jp.co.hateblo.bomberhead} & 11 & 3.88 & 1.12 & 4 & 0.69 & 0.13 \\
\textbf{com.guidoo.lulo.booxx} & 17 & 5.69 & 1.69 & 7 & 1.56 & 0.38 \\
\textbf{ch.rmy.android.http} & 10 & 2.50 & 0.88 & 8 & 1.81 & 0.37 \\
\hline
\end{tabular}}
\label{tab:task3}
\end{table}

\begin{table}
\centering
\caption{Feedback from app developers.
}
\scalebox{0.7}{\begin{tabular}{c|c|c|c|c}
\hline
\textbf{App} & \textbf{\# Star} & \textbf{Version} & \textbf{IssueID} & \textbf{Issue State}\\
\hline
\textbf{OpenPods} & 613 & v1.7 (18) & \#142 & \textbf{Merged} \\ \hline
\textbf{motioneye-client} & 27 & v1.0.0-alp9 (10000008) & \#22 & \textbf{Merged} \\ \hline
\textbf{TapUnlock} & 28 & v2.1.0 beta (13) & \#5 & \textbf{Merged} \\ \hline
\textbf{ItsDicey} & 4 & v1.0.1 (2) & \#3 & \textbf{Merged} \\ \hline
\textbf{greentooth} & 20 & v1.12 (5) & \#4 & \textbf{Merged} \\ \hline
\textbf{knightsofalentejo} & 8 & v5.0.0-RC-1 (5021) & \#6 & \textbf{Merged} \\ \hline
\textbf{mundraub-android} & 32 & v1.236 (237) & \#326 & \textbf{Merged} \\ \hline
\textbf{DriSMo} & 23 & v1.0.3 (17) & \#4 & \textbf{Merged} \\ \hline
\textbf{Icicle for Freenet} & 10 & v0.4 (4) & \#5 & \textbf{Merged} \\ \hline
\textbf{Anki-Android} & 5.4k & v2.15.6 (21506300) & \#10472 & Positive Response \\ \hline
\textbf{ActivityManager} & 69 & v4.2.0 (415) & \#6 & Positive Response \\ \hline
\textbf{Mensa} & 9 & v1.7.0 (38) & \#5 & Positive Response \\ \hline
\textbf{mundraub-android} & 32 & v1.236 (237) & \#326 & Positive Response \\ \hline
\end{tabular}}
\label{tab:Feedback}
\end{table}

\Cref{fig:distribution} shows the distribution of participants' score results on the repair effect of \tool under the condition of known issues (both participants' scores are greater than 3). 
{More than one-third of the results (36.6\%) are full scores (i.e., 5), and the results higher than 4 accounts for 81.3\% of all scores. The final calculated average score is {4.218}.}
Most of the participants said that the color contrast of the repaired UI was improved and easy to read, the replacement color selected during repair is also more coordinated with the style of the original UI, and the feeling of the page is better than the original UIs. In addition, some participants said that the repair of \tool is more effective for people with low vision, and the color matching of the original UI also has practical advantages under the condition of normal vision.

% \sen{Task 3.}
Table~\ref{tab:task3} shows the feedback results of users on the overall coordination of the repaired apps at the app level. $C_{All}^{Re}$ and $P_{All}^{Re}$ respectively represent the number of components and UI pages repaired by \tool, while $C_{All}^{Un}$ and $P_{All}^{Un}$ respectively represent the average number of components and UI pages that are considered uncoordinated by the user. While $C_{Re}^{Un}$ and $P_{Re}^{Un}$ are the number of components repaired by \tool and contained in the components or UI pages circled by users (considered uncoordinated with the design style of the original app). It can be seen that the components and UI pages repaired by \tool are rarely considered to be uncoordinated with the original app, which also indicates that when \tool is applied to an app with color-related issues including multiple UI interfaces, the color selected by \tool is consistent with the original app design.

\subsubsection{\textbf{Feedback from app developers}}

\noindent \textbf{Setup.}
{To understand the views of app developers, we randomly selected 40 open-source apps containing color-related accessibility issues from F-Droid~\cite{fdroid} and used \tool to repair them, {aiming to obtain feedback from them about our issue repair results}.}
{Although there are many issues detected and repaired in each app, to investigate how different app developers of specific apps act towards accessibility issues and the repair result, we randomly selected one issue from each app and submit pull requests in the corresponding GitHub repositories, in which we introduced the focus of \tool, implementation functions, and the repair results to developers. We also asked for comments about \tool, the repair result, and suggestions for improvement.}

\begin{figure}
\centering
\includegraphics[width=0.5\textwidth]{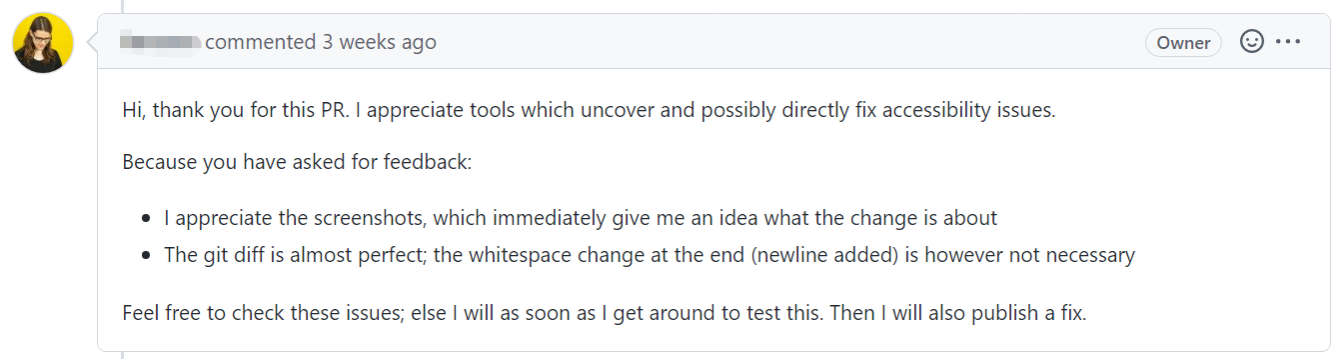}
 \caption{Feedback from the real app developers.}
\label{fig:feedback}
\end{figure}

\smallskip
\noindent \textbf{Result.}
As shown in Table~\ref{tab:Feedback}, till now, we received 9 merged pull requests and 4 positive comments. Meanwhile, we display the app name, app version, the number of stars, and the id of the pull requests in this table. 
{During the traceability analysis, the developers claimed ``\textit{That looks interesting}''~\cite{Feedback1} and ``\textit{Good you made a PR and bring in your knowledge}''~\cite{Feedback2} in their feedback, indicating the usefulness and practicability of \tool.}
\Cref{fig:feedback} shows an example of feedback from an app developer.
{She pays attention to repairing such issues and hopes to have tools that uncover and directly fix accessibility issues.}
{Although we find that some developers even do not know about these contrast issues~\cite{di2022making}, they have realized the importance of accessibility through \tool and want to use it to automatically repair their apps, which shows that \tool is meaningful. At the same time, other developers think that their limitation is the time to implement the fixes and worry about the difficulty of positioning.}
{They also hope to have a tool that can automatically analyze relevant issues, and \tool just implements this function and realizes the positioning of related components and automated recommendation of replacement colors, which shows that \tool has practical significance.}
{More importantly, \tool can clearly make more developers aware of these issues, so they can effectively avoid them during development.}

\section{Discussion}
\subsection{\textbf{Limitations}}
\subsubsection{Scope-to-repair.}
\textit{(1)} \tool repairs the apps based on the input test results, so we only repair the problematic components raised in the test report until now. If there are other components with color-related accessibility issues, but they are not detected by the detection tool, \tool will not repair them. In addition, most of the repair failures are due to the limitations of \textit{bounds} shown in our experiments in~\S~\ref{subsec:rq1}. Although using \textit{bounds} can effectively solve the problem of text contrast, it cannot accurately locate the problem of image contrast. So the repair rate of image contrast is slightly lower. But if the information identifying the components in the detection report is more accurate, the less this restriction will be. In fact, although the repair of the apps highly depends on the detection report, if the other effective detection tools can be integrated in the future, \tool also works for them because it is scalable in practice.
\subsubsection{Object-to-repair.} 
\textit{(2)} \tool adopts the method of static analysis in the localization and repair stage to analyze and extract attributes from the app UI layout file and its resource file. However, the properties of some components are set in the app source code. Currently, we can only handle the basic implementation by {Java or Kotlin} code such as using official API like \textit{setTextColor}, as shown in~\Cref{listing:implementation}. We noticed that some third-party libraries provide new API interfaces to restructure the official APIs. \tool cannot resolve the restructured implementations, but they actually do not belong to the research scope of our work. {Similarly, Jetpack uses a new view structure to implement GUIs, therefore, \tool has restrictions on apps using Jetpack. Based on our experiments and investigation, most apps use the traditional Android XML layouts (i.e., Android View) to design and implement their UI pages. Therefore, the current version of \tool is applicable to most recent apps.}

\subsection{\textbf{Threats to validity}}
As we mentioned in~\S~\ref{subsec:rq2}, the proportion of the first strategy directly depends on the size of the color reference DB we constructed in advance. If the size changes, the proportions of the first strategy will change accordingly. We are continuing to expand the reference DB by running more apps. In fact, under the current situation, the proportion of the first strategy is already the largest, that is, it has the greatest impact on the color selection results.

\balance
\section{Related Work}
\subsection{Accessibility Issue Repair}
\subsubsection{Mobile platform}
For mobile accessibility, automated repair has been a fresh research direction in recent years. Most of the existing studies focused on the issue categories with high proportions such as the issues of \textit{item label and touch target}. As for the \textit{item label} issues, the goal of repair is to augment or predict the missing content labels for UI components. For example, Zhang et al.~\cite{zhang2018robust} developed methods for robust annotation of app interface elements by leveraging social annotation techniques that have been used on the web. Chen et al.~\cite{chen2020unblind} trained a deep learning model named {LabelDroid} to predict the missing labels. {COALA}~\cite{mehralian2021data} was a similar work based on using deep learning algorithms. Moreover, crowd-sourcing techniques are also used to recommend the labels of UI components in~\cite{brady2015crowdsourcing}. In terms of \textit{touch target} issues, named size-based accessibility issues, Alotaibi et al.~\cite{alotaibi2021automated} leveraged a genetic algorithm guided by a fitness function to automatically repair them. {All above studies focus on addressing one specific type of issue due to the diverse characteristics of different issues.}

Compared with these studies, we focus on the other issue categories including text contrast and image contrast, named color-related accessibility issues. However, this category is of great importance and critical not only for its high proportion but also for the impact of mobile accessibility.

\subsubsection{Web platform}
Substantial efforts are put into automatically fixing accessibility problems in web settings~\cite{mahajan2017xfix,mahajan2018automated,mahajan2017automated,mahajan2018automated1,panchekha2016automated}, but some of them focused more on Mobile Friendly Problems, 
which can inspire the repair of size-related issues, however, cannot benefit color-based accessibility issues. 
In the works related to color-related accessibility issues, the re-coloring tool~\cite{flatla2013sprweb} enhanced the accessibility of the entire page by changing the color matching of the entire web page. Tools~\cite{sandnes2021inverse,hansen2019still,richardson2014color} that modify the color matching of some web components with accessibility issues also tried to improve the contrast of components by changing the problematic text background color pairs, but they all lack consideration of the overall design style of the original web page.
Last but not least, the implementation mechanisms are significantly different for web apps and Android apps, which directly distinguish repair solutions. 

\subsection{\textbf{Accessibility Issue Detection and Analysis}}
%\noindent{\textbf{Accessibility Issue Detection}.}
%\subsubsection{Mobile platform}
Compared with mobile app testing including functional testing~\cite{fan2018large,fan2018efficiently,su2020my,yang2023compatibility} and security testing~\cite{chen2018mobile,chen2020empirical,chen2022ausera}, mobile app accessibility testing is studied to a lesser extent. Silve et al.~\cite{silva2018survey} surveyed the available approaches for detecting accessibility issues. The existing approaches can be categorized into static and dynamic methods. Android Lint~\cite{lint} can report missing content labels, but it is ineffective for other issue categories. Accessibility Scanner~\cite{scanner} can detect issues with the help of manual exploration of the app but is limited to the low activity coverage. To mitigate such problems, Eler et al.~\cite{eler2018automated} developed a model, named {MATE}, by generating testing cases specifically for accessibility testing. Similarly, Salehnamadi et al. proposed Latte~\cite{salehnamadi2021latte} and Groundhog~\cite{salehnamadi2022groundhog} by reusing tests written by developers or automatically generated to validate the accessibility of those use cases. Recently, Alshayban et al.~\cite{alshayban2020accessibility} detected issues by deploying Monkey~\cite{monkey}. However, in a recent work, Chen et al.~\cite{chen2021accessible} identified such a solution is not effective and they further proposed Xbot to automatically detect accessibility issues by leveraging app instrumentation and data-flow analysis.
Recently, several works~\cite{alotaibi2022automated, mehralian2022too} have also focused on the interactive accessibility of apps when users with disabilities are using Assistive Technologies, such as TalkBack~\cite{talkback} and VoiceOver~\cite{voiceover}.

A large number of empirical studies focused on investigating the characteristics of mobile accessibility. Ross et al.~\cite{ross2018examining} unveiled some common labeling issues. Their following work~\cite{ross2020epidemiology} studied the properties of each accessibility type. Yan et al.~\cite{yan2019current} investigated if the apps violate the accessibility guidelines and further introduced the degree of violation. Vendome et al.~\cite{vendome2019can} proposed a taxonomy in terms of the aspects of accessibility issues by analyzing the developers' posts on Stack Overflow. Alshaybana et al.~\cite{alshayban2020accessibility} proposed a metric named \textit{inaccessibility issue rate} to measure the distribution of such a metric for each app, each issue type, and app categories. Moreover, they also observed this research field from app developers and users. Chen et al.~\cite{chen2021accessible} also conducted a large-scale empirical study from the perspective of the issues themselves and revealed many fine-grained findings.

\section{Conclusion}
In this paper, we propose \tool to automatically repair the color-related accessibility issues for Android apps. Our approach builds a large-scale reference DB to help design {a context-aware} color selection technique as well as well-defined criteria and an effective attribute-to-repair localization method. Based on these key phrases, \tool can identify the optimal color replacement for automated repair and further generate a new repackaged APK for app developers. Our comprehensive experiments including a user study demonstrate the effectiveness, efficiency, and usefulness of our approach from different aspects. We finally highlight that the feedback from several real app developers is quite positive and the merged pull requests on GitHub confirm the practicality of \tool.

\section{Data Availability}
We have released the code of \tool on GitHub~\cite{ourgithub}, the constructed reference DB based on 9,978 apps and the 100 APK files used in our experiment on Google Drive~\cite{100apks}. To facilitate developers to understand the repair performance, we also have uploaded the data of the user study and some other repair cases to our website~\cite{ourwebsite}.

\begin{acks}
This work was supported by the National Natural Science Foundation of China (No. 62102284, 62102197).
\end{acks}

\clearpage
\balance
\bibliographystyle{ACM-Reference-Format}
\bibliography{acmart.bib}

\end{document}